\documentstyle[11pt]{article}
\newcommand{\be}{\begin{equation}}
\newcommand{\ee}{\end{equation}}
\newcommand{\ba}{\begin{eqnarray}}
\newcommand{\ea}{\end{eqnarray}}
\newcommand{\bea}{\begin{eqnarray}}
\newcommand{\eea}{\end{eqnarray}}

\begin{document}
\title{Extending the four-body problem of   Wolfes to non-translationally
invariant interactions} 

\author{ A. Bachkhaznadji \\
Laboratoire de Physique Th\'eorique \\
D\'epartement de Physique \\
Universit\'e  Mentouri \\
Constantine, Algeria\\
\\ 
M. Lassaut    \\
Institut de Physique Nucl\'eaire\\
 IN2P3-CNRS and Universit\'e  Paris-Sud,   \\
F-91406 Orsay CEDEX,  France \\ [3mm]
\date{\today }}
\maketitle

Abstract :

We propose and solve exactly the Schr\"odinger equation  of a bound quantum 
system consisting in four particles moving on a real line with both translationally invariant four 
particles interactions of Wolfes type \cite{Wolf74} and additional non  translationally
invariant four-body potentials. 
We also  generalize and solve exactly this problem in  any $D$-dimensional space by providing
full eigensolutions and the corresponding energy spectrum.
 We discuss the domain of the coupling constant where the irregular solutions becomes physically 
acceptable


\newpage

\section{Introduction}
There exists a very limited number of exactly solvable many-body systems,
even in one dimension space (1D) \cite{mattis,sutherland}.
The Calogero model constitutes  one of the famous ones,  which was
exhaustively studied \cite{Calo69p,Calo71}.
A survey of many quantum integrable systems was done by
Olshanetsky and Perelomov  \cite{Perelomov1983}.
They classified the systems with respect to Lie algebras.
Point interactions have also been considered, still in $D=1$ 
\cite{albe1,albe2}.

The quest for exactly solvable non trivial quantum  problems
of few interacting particles on the line or on the circle still retains
attention.
Early works of three-body linear problems of
 Calogero-Marchioro-Wolfes  \cite{Calo69,CM74,Wolfes74}
have been followed by new extensions and cases.  In a non exhaustive way, we
quote, for instance the three-body version of the Sutherland problem, with only
a translationally invariant
three-body potential, solved by Quesne   \cite{ques2}. By using supersymmetric
quantum mechanics Khare {\it et al.} \cite{KB}\, gave examples of algebraically
solvable three-body problems of Calogero type on the line with additional
translationally invariant two-and/or three-body potentials.

A new integrable model  on the line of the Calogero type  with a
non-translationally invariant two-body potential, was worked out by Diaf {\it et
al.}  \cite{DKML}. The extension of this linear model  to $D$-dimensional space was done in
\cite{BLL2007}. We note also that a generalization of this latter linear model
was solved by Meljanac  {\it et al.} \cite{Meljanac2007}, by emphasizing the
underlying conformal  $SU(1,1)$ symmetry of the model. \\
Recently, some exactly solvable generalizations of the Calogero  \cite{Calo69}
and the Calogero-Marchioro-Wolfes three-body linear problems
\cite{CM74,Wolfes74}, have been proposed by Bachkhaznadji {\it et
al.} \cite{BLL2009}, with the  introduction of   non-translationally invariant
three-body potentials. \\
Finally, we can cite the work of Haschke and  R\"uhl \cite{Has98} concerning  the
construction of exactly solvable quantum models of Calogero and Sutherland type
with translationally invariant two-and four particles interactions.

The purpose of this  paper is to study a completely solvable four-body quantum
problem  by providing explicitly the eigenvalues  and the complete set of
associated eigensolutions
of the considered Schr\"odinger equation. This is possible  through an
appropriate coordinates transformation.
 We consider   four particles bounded in an harmonic trap
moving on the line with only four-particles interactions.  One of these is a
translationally invariant potential,
which was introduced by Wolfes  \cite{Wolf74}.
The other interactions are   non-translationally invariant four-body potentials.
\\
 The irregular solutions of the problem are also studied, when they  become
square integrable, and thus physically acceptable. Such a situation occurs for a
suitable domain of the coupling constant. \\
  This model can be extended to any  $D$-dimensional space, and
 solved exactly by deriving the full expressions of both the energy spectrum
and the eigensolutions.

The paper is organized   as follows.
In section {\bf 2} we expose and solve the problem
for the linear  case. The section {\bf 3} is devoted to 
extension to $D$-dimensional problem.  Our conclusions are drawn in section {\bf
4}.

\section{A generalization of the linear Wolfes four-body problem}

We consider the four-body Hamiltonian on the line

\begin{equation}
H=\sum_{i=1}^{4}\left( -\frac{\partial ^{2}}{\partial x_{i}^{2}}
+\omega ^{2}x_{i}^{2}\right) +2 \lambda  \sum_{i \ne j \ne k \ne m} \frac{1}{
(x_{i}+x_j-x_k-x_m)^2}+\frac{4 \mu }{(\sum_{i=1}^{4}x_{i})^2}+\frac{ \beta}{\sum_{i=1}^4
 x_i^2}
\label{3c-1}
\end{equation}
or more explicitly
\begin{eqnarray}
H &=&-\frac{\partial ^{2}}{\partial x_{1}^{2}}-\frac{\partial ^{2}}{\partial
x_{2}^{2}}-\frac{\partial ^{2}}{\partial x_{3}^{2}}-\frac{\partial ^{2}}{\partial x_{4}^{2}}
  +\omega^{2}(x_{1}^{2}+x_{2}^{2}+x_{3}^{2}+x_4^2)+
  \frac{4 \mu }{(x_{1}+x_{2}+x_{3}+x_4)^2}  \nonumber\\
&&+ 4 \lambda \left[ \frac{1}{(x_{1}+x_{2}-x_3-x_4)^{2}}+\frac{1}{(x_{1}+x_{3}-x_2-x_4)^{2}} 
+\frac{1}{(x_1+x_4-x_{2}-x_{3})^{2}}\right] \nonumber\\
& +& \frac{\beta}{x_{1}^{2}+x_{2}^{2}+x_{3}^{2}+x_4^2}.
\label{H1}
\end{eqnarray}

This Hamiltonian represents a system of four particles on the line with the same mass
(with units $\hbar=2 m=1$) interacting via only  four-body  potentials.
 One potential is translationally  invariant with coupling constant $\lambda$ of Wolfes type \cite{Wolf74}
and the two others with    coupling constants $\mu,\beta$ are not    translationally  invariant.
 The whole system is confined in an harmonic oscillator trap.
 
The problem is solved in the following way. Setting
\bea
R & = & \frac{x_1+x_2+x_3+x_4}{2} \nonumber\\
s & = & \frac{x_1+x_2-x_3-x_4}{2} \nonumber\\
t & = &  \frac{x_1+x_3-x_2-x_4}{2} \nonumber\\
u & = &  \frac{x_1+x_4-x_2-x_3}{2} 
\label{coord}
\eea
the Hamiltonian, Eq.(\ref{H1}), becomes 

\begin{eqnarray}
H &=&-\frac{\partial ^{2}}{\partial R^2}-\frac{\partial ^{2}}{\partial
s^{2}}-\frac{\partial ^{2}}{\partial t^{2}}-\frac{\partial ^{2}}{\partial u^{2}}
  +\omega^{2}(R^2+ s^2+ t^2+u^2) 
 \nonumber\\
&& + \frac{ \mu }{R^2} + \lambda \left( \frac{1}{s^{2}}+\frac{1}{t^{2}} +  
\frac{1}{u^{2}}\right) +\frac{\beta}{R^2+ s^2+ t^2+u^2}.
\label{H2}
\end{eqnarray}

Note that, if $\beta=0$, the problem is separable and the derivation of the solutions
is straightforward. Otherwise, it is not  separable in  $\{R,s,t,u\}$ variables.
To overcome this situation we introduce the following  hyperspherical coordinates: 
\begin{eqnarray}
R &=&r\cos \alpha, \quad s=r\sin \alpha \cos \theta, \quad t=r\sin \alpha \sin \theta \sin \varphi,
\quad u=r\sin \alpha \sin \theta \cos \varphi ,\quad  \nonumber \\
0 &\leq &r<\infty ,\quad \quad  0\leq \alpha \leq \pi ,\quad\quad\quad\quad   0\leq \theta \leq \pi,\quad \quad\quad\quad\quad 0\leq
\varphi \leq 2\pi .  \label{3c-18bis}
\end{eqnarray}
The stationary Schr\"{o}dinger equation is then written  as:
\begin{eqnarray}
&&\left\{ -\frac{\partial ^{2}}{\partial r^{2}}-\frac{3}{r}\frac{\partial }{
\partial r}+\omega ^{2}r^{2}  +\frac{\beta}{r^2} +  \frac{1}{r^{2}}\left[
 -\frac{\partial ^{2}}{\partial \alpha ^{2}}   -2 \cot \alpha \frac{\partial }{\partial \alpha } + \frac{\mu }{\cos^2 \alpha}
 \right.\right. \nonumber\\
&+& \left.\left. 
\frac{1}{\sin ^{2}\alpha}  \left(  -\frac{\partial ^{2}}{\partial \theta ^{2}}-\cot \theta \frac{\partial }{\partial \theta }
+\frac{\lambda}{ \cos ^{2}\theta }
 \right.\right. \right. \nonumber\\
 & & \left.\left.\left. + \frac{1}{\sin^{2}\theta }\left(- 
\frac{\partial ^{2}}{\partial \varphi ^{2}}+ \frac{4 \lambda}{\sin^{2}2 \varphi }
\right) \right) \right] \right\} \Psi(r,\alpha,\theta,\varphi)  =E\Psi(r,\alpha,\theta,\varphi)   ,  \label{3c-49}
\end{eqnarray}
where $\Psi (r,\alpha,\theta ,\varphi )$ represent the eigensolutions associated to eigenenergy $E$.

This four-body problem described by  this  equation  (\ref{3c-49}) may be mapped to
 the problem of one particle in four dimensional space, with a non central potential of the form
\begin{equation}
V(r,\alpha,\theta ,\varphi )=f_{1}(r)+\frac{1}{r^2} \left( f_2(\alpha) +\frac{1}{\sin ^{2}\alpha } \left[
f_{3}(\theta )+ \frac{f_{4}(\varphi )}{\sin ^{2}\theta } \right] \right).
\end{equation}
It is then clear that the problem becomes separable in the  four variables $\{r,\alpha,\theta,\varphi\}$.
To find the solution we factorize the wave function as follows:
\begin{equation}
\Psi_{k,\ell,m,n} (r,\alpha,\theta ,\varphi )=\frac{F_{k,\ell,m,n}(r)}{r \sqrt{r}} \frac{G_{\ell,m,n}(\alpha)}{\sin \alpha}
 \frac{\Theta_{m,n} (\theta )}{\sqrt{\sin
\theta }}\Phi_n (\varphi ).  \label{3c-6} 
\end{equation}
Accordingly,  equation (\ref{3c-49}) separates in  four   decoupled  differential equations:
\begin{equation}
\left( -\frac{d^{2}}{d\varphi ^{2}}+\frac{ 4 \lambda }{\sin^{2} 2\varphi }
\right) \Phi_n (\varphi )=B_n\Phi_n (\varphi ),\quad  \label{3c-7}
\end{equation}
\begin{equation}
\left( -\frac{d^{2}}{d\theta ^{2}}+\frac{(B_n-\frac{1}{4})}{\sin ^{2}\theta }
+\frac{\lambda}{\cos^2\theta} \right) \Theta_{m,n} (\theta )=C_{m,n} \Theta_{m,n} (\theta ),\qquad  \label{3c-8}
\end{equation}
\begin{equation}
\left( -\frac{d^{2}}{d\alpha ^{2}}+\frac{C_{m,n}-\frac{1}{4}}{\sin ^{2}\alpha }
+\frac{\mu}{\cos^2 \alpha} \right) G_{\ell,m,n} (\alpha )=D_{\ell,m,n} G_{\ell,m,n} (\alpha ),\qquad  \label{3c-9}
\end{equation}
and 
\begin{equation}
\left( -\frac{d^{2}}{dr^{2}}+\omega ^{2}r^{2}+\frac{\beta +D_{\ell,m,n}-\frac{1}{4}}{r^{2}
}\right) F_{k,\ell,m,n}(r)=E_{k,\ell,m,n}\  F_{k,\ell,m,n}(r) \ .  \label{3c-10}
\end{equation}
 
In the interval $0 \leq \varphi \leq 2 \pi$ the potential involved in  equation (\ref{3c-7}) has a periodicity of $\pi$ 
and has singularities at $\varphi=k \pi/2,k=0,1,2,3$. The equation is first solved in the interval $[0,\pi/2]$.
 It can be treated if and only if $\lambda >-1/4$ otherwise the operator has several self-adjoint extensions,
each of them leading to a different spectrum \cite{Zno,Barry}. 
The regular eigensolutions of equation (\ref{3c-7}) on $[0,\pi/2]$ read  \cite{Perelomov1983,JZ}
\be  
 \Phi_n (\varphi )= (\sin 2 \varphi)^{1/2+a} \  C_n^{(1/2+a)}(\cos2 \varphi)
\ee
where the  $C_n^{(1/2+a)}$ denote the Gegenbauer Polynomials \cite{erd}. The  corresponding  eigenvalues 
are
\be  
B_n=4 \left(\frac{1}{2} +a+n \right)^2 
\label{bn}
\ee
with 
\be  
 n=0,1,2,... \qquad\quad a =\frac{\sqrt{1+4 \lambda}}{2} 
\label{symba}
\ee 
The extension  to the whole interval $[0,2 \pi] $ is made in two steps. First one, from $[0,\pi]$ up to 
$[\pi,2\pi]$ using the periodicity of the solutions.
On the other hand, the symmetric extension of the solutions in the interval $[\pi/2,\pi]$ reads 
\be  
 \Phi_n (\varphi )= (-\sin 2 \varphi)^{1/2+a} \  C_n^{(1/2+a)}(\cos2 \varphi)
\ee
 so that  the real power of $ (-\sin 2 \varphi)$ is defined. In a compact form we obtain
\bea 
   \Phi_n (\varphi ) &= &  (\epsilon_1 \sin 2 \varphi)^{1/2+a} \  C_n^{(1/2+a)}(\cos2 \varphi) 
\label{phin}  \\
  & & \epsilon_1=\pm 1, \qquad\quad \frac{1-\epsilon_1}{2} \frac{\pi}{2} \leq \varphi \leq  \frac{3-\epsilon_1}{2} \frac{\pi}{2} \nonumber
\eea
Note  that when $a=1/2, (\lambda=0)$, a $\delta$-pathology occurs at $\varphi=\pi/2$.

On the other hand,  the antisymmetric extension of the solution reads 
\bea 
   \Phi_n (\varphi ) &= & {\rm sgn}(\sin(2 \varphi)) (\epsilon_1 \sin 2 \varphi)^{1/2+a} \  C_n^{(1/2+a)}(\cos2 \varphi) 
\label{phina}  \\
  & & \epsilon_1=\pm 1, \qquad\quad \frac{1-\epsilon_1}{2} \frac{\pi}{2} \leq \varphi \leq  \frac{3-\epsilon_1}{2} \frac{\pi}{2} \nonumber
\eea
where ${\rm sgn}(x)=x/|x|, x \ne 0$, denotes the sign  of the variable $x$.

 Generally, only the regular  solution, $\Phi _{n}^{(+)}, $ corresponding  to $1/2+a$, is retained.
However,  the irregular solution,  $\Phi _{n}^{(-)} $, which is distinct from $\Phi _{n}^{(+)} $ for
$\lambda >-1/4$, corresponding to $1/2-a$ is physically  acceptable when the  Dirichlet condition is satisfied for $-1/4 < \lambda \leq 0$ (attractive potentials). 
 If we release the Dirichlet condition, and ask only for the square integrability of the solution,  as in \cite{Murthy92}, 
then $\Phi _{n}^{(-)}$ can be retained for $-1/4<\lambda <3/4$ \cite{BLL2009}.
For the irregular antisymmetric  solution,  a (derivative of) $\delta$-pathology occurs at $\varphi=\pi/2$, when $a=1/2, (\lambda=0)$.

Introducing the parameter
\be   
 b_n=\sqrt{B_n}=1+2 a + 2 n
\label{bnp}
\ee
(where  we have only considered the positive root $b_n=\sqrt{B_n}$ because the other root $b_n=-\sqrt{B_n}$
leads to non-square integrable solutions for most values of $n$) 
the solution of Eq.(\ref{3c-8}) on the whole interval $[0,\pi]$ can be  obtained  as  \cite{Perelomov1983,BLL2009}
\bea   
 \Theta_{m,n}(\theta)&=& {\rm sgn(\cos  \theta)}^{s_{\theta}}  (\sin \theta)^{b_n+1/2} \ (\epsilon_2  \cos  \theta)^{c+1/2} \ P_m^{(b_n,c)}(\cos 2 \theta)
\label{thetap} \\
  & & m=0,1,2,... \qquad\quad c =\frac{1}{2} \ \sqrt{1 + 4 \lambda} \label{symbc} \\
  & &   \epsilon_2 =\pm 1, \qquad\quad  \frac{1-\epsilon_2}{2} \frac{\pi}{2} \leq \theta \leq  \frac{3-\epsilon_2}{2} \frac{\pi}{2}
\nonumber
\eea
corresponding to the eigenvalue
\be   
C_{m,n}= (2 m + b_n + c + 1)^2  \ .
\label{Cmn}
\ee
In fact, $a=c$, if one takes the definitions Eqs.(\ref{symba},\ref{symbc}).
In Eq.(\ref{thetap}), the $P_m^{(b_n,c)}$ denote the Jacobi polynomials \cite{erd}.
The value $s_{\theta}=0,1$ according to the fact that the solution has been extended in a symmetric (antisymmetric) way
from $\theta \in [0,\pi/2]$ to  $\theta \in [\pi/2,\pi]$.
Note  that, when $c=1/2, (\lambda=0)$, a $\delta$-pathology occurs for the symmetric solution at $\theta=\pi/2$.

The equation (\ref{3c-9}) is solved in the same manner as  for equation  (\ref{3c-8}).
Setting 
\be  
c_{m,n}=\sqrt{C_{m,n} \ }=2 m + b_n + c + 1
\label{cmn}
\ee

(where  we have only considered the positive root $c_{m,n}=\sqrt{C_{m,n}}$ because the other root $c_{m,n}=-\sqrt{C_{m,n}} $
leads to non-integrable solutions for most values of $m$)
the solution writes as

\bea 
 G_{\ell,m,n} (\alpha ) & = &{\rm sgn(\cos  \alpha)}^{s_{\alpha}}   (\sin \alpha)^{c_{m,n}+1/2} \ (\epsilon_3  \cos  \alpha)^{d+1/2} \
  P_{\ell}^{(c_{m,n},d)}(\cos 2 \alpha) \label{Glmn} \\
   & & \ell=0,1,2,... \qquad\quad d =\frac{1}{2} \ \sqrt{1 + 4 \mu}  \label{dd} \\
  & & \epsilon_3 =\pm 1 \qquad\quad  \frac{1-\epsilon_3}{2} \frac{\pi}{2} \leq \alpha \leq  \frac{3-\epsilon_3}{2} \frac{\pi}{2} \nonumber
\eea
corresponding to the eigenvalue
\be   
D_{\ell,m,n} = (2 \ell +c_{m,n}  +d + 1)^2  \ .
\label{3c-31}
\ee 
The value $s_{\alpha}=0,1$ according to the fact that the solution has been extended in a symmetric (antisymmetric) way
from $\alpha \in [0,\pi/2]$ to  $\alpha \in [\pi/2,\pi]$.
Note  that when $d=1/2, (\mu=0)$, for symmetric solutions, a $\delta$-pathology occurs at $\alpha=\pi/2$.

On the other hand, the reduced radial equation (\ref{3c-10})
is solved in the interval  $0\leq r<\infty ,$ with the condition of square integrability for the solutions.
It implies $F_{k,\ell,m,n}(r) \to 0$ as $r \to \infty$.

We have to impose $\beta + D_{\ell,m,n} >0$ in order to treat the centrifugal barrier in the vicinity of $r=0$.
Note that taking $\beta + D_{\ell,m,n} =0$ leads to several self-adjoint extensions parametrized by a phase.
This fact has been discussed in \cite{BLL2009}. More details can be found in \cite{basu,giri}.
Also attractive barriers  $\beta + D_{\ell,m,n} <0$,  mentioned in \cite{BLL2009}, have been treated in \cite{case,gupta,camblong}.
Taking into account the definition of $D_{\ell,m,n}$,  Eq.(\ref{3c-31}), we must have
\begin{equation}
\beta +D_{\ell,m,n}=\beta +\left(2 \ell+2 m + 2 n + 2 a + c + d + 3 \right)^{2}> 0, \quad
\forall n\geq 0,\quad \forall m\geq 0, \quad \forall \ell \geq 0.  \label{3c-32bisp}
\end{equation}
For positive values $a,c,d$,  the  quantity $\beta +D_{\ell,m,n}$ is minimal for  $n=0,m=0,\ell=0$ and $a=c=d=0$ ( recall that $a \geq 0$, see (\ref{symba}),
 that $c \geq 0$, see (\ref{thetap}) and that $d \geq 0$  see (\ref{dd})). 
It puts constraint on  $\beta$ to satisfy  $\beta >-9$ when $a$,$c$ and $d$ are positive.
 We introduce the auxiliary parameter  $\kappa _{\ell,m,n}$ defined by
\begin{equation}
\kappa _{\ell,m,n}^{2}=\beta +D_{\ell,m,n}, \qquad \kappa _{\ell,m,n}=
\sqrt{\beta +D_{\ell,m,n} \ }.
\label{3c-33}
\end{equation}
 The solution of the radial equation (\ref{3c-10})  is \cite{BLL2009} 
\begin{equation}
F_{k,\ell,m,n}(r)=r^{\kappa _{\ell,m,n}+\frac{1}{2}}\exp \left(-\frac{\omega r^{2}}{2}
\right)L_{k}^{(\kappa _{\ell,m,n})}(\omega r^{2}),\qquad k=0,1,2...,  \label{3c-37}
\end{equation}
and it is  associated to the eigenenergy
\begin{equation}
E_{k,\ell,m,n}=2\omega (2k+\kappa _{\ell,m,n}+1),\qquad k=0,1,2....  \label{3c-38}
\end{equation}
The $L_k^{(q)}$ are the generalized Laguerre polynomials \cite{erd}.
  
Taking into account all  information, we conclude  that the physically acceptable solutions of 
the Schr\"odinger equation  (\ref{3c-49}) are 
\begin{eqnarray}
\Psi _{k,\ell,m,n}(r,\alpha,\theta ,\varphi ) &=&r^{\sqrt{\beta +(2 \ell+ 2m +2n+2a+c+d+3)^{2} \ }-
1}e^{-\frac{\omega r^{2}}{2}}L_{k}^{\left( \sqrt{\beta +(2 \ell+ 2m +2n+2a+c+d+3)^{2}  \ }
\right) }(\omega r^{2})  \nonumber \\
&&\times   {\rm sgn(\cos  \alpha)}^{s_{\alpha}}    (\sin \alpha )^{2m +2n+2 a+c+\frac{3}{2}} \ (\epsilon_3  \cos \alpha )^{d+1/2} 
P_{\ell}^{(2 m + 2 n + 2 a + c +2,d)}(\cos 2 \alpha)
\nonumber\\
&&\times   {\rm sgn(\cos  \theta)}^{s_{\theta}}    (\sin \theta )^{2 n+2a+1} (\epsilon_2 \cos \theta)^{c+\frac{1}{2}} 
P_{m}^{\left( 2n + 2 a + 1,c \right)}(\cos 2 \theta ) 
\nonumber\\
&&\times    {\rm sgn(\sin(2 \varphi))}^{s_{2 \varphi}}  (\epsilon_1 \sin 2\varphi )^{a+\frac{1}{2}} C_{n}^{\left( a+\frac{1}{2}\right)
}(\cos 2 \varphi ),\,  \label{3c-99p} \\
k &=&0,1,2,...,\qquad \ell=0,1,2,...,\qquad m=0,1,2,...,,\qquad n=0,1,2,...,   \nonumber \\
& & \frac{1-\epsilon_1}{2} \frac{\pi}{2}  \leq \varphi  \leq \frac{3-\epsilon_1}{2} \frac{\pi }{2} , \epsilon_1=\pm 1, \nonumber\\
& &  \frac{1-\epsilon_2}{2} \frac{\pi}{2}  \leq \theta \leq \frac{3-\epsilon_2}{2} \frac{\pi }{2} , \epsilon_2=\pm 1,
\quad \frac{1-\epsilon_3}{2} \frac{\pi}{2}  \leq \alpha \leq 
\frac{3-\epsilon_3}{2} \frac{\pi }{2} , \epsilon_3=\pm 1 \nonumber\\
& & \quad a=\frac{1}{2}\sqrt{
1+4 \lambda } \ ,\quad c=\frac{1}{2}\sqrt{1 + 4 \lambda  \ } \quad d=\frac{1}{2}\sqrt{1 + 4 \mu \ } \ .
 \nonumber    
\end{eqnarray}
Here  $s_{2 \varphi}=0,1$ according to the parity of the solution for $\varphi \in [0,\pi]$.
We note that $\delta$-pathologies occur for respectively $d=1/2$, $a=c=1/2$,
in  Eq.\ref{3c-99p}, when symmetric solutions ( i.e., $s_{\alpha}$, $s_{\theta}$ or $s_{2 \varphi}$
 equal to zero) are considered.

The normalization constants $N_{k,\ell,m,n}$ can be  calculated from
\begin{eqnarray}
&& \int_{0}^{+\infty }r^{3}dr \ \int_{(1-\epsilon_3) \pi/4}^{(3-\epsilon_3) \pi/4} \sin^2(\alpha) \ d\alpha 
\int_{(1-\epsilon_2) \pi/4}^{(3-\epsilon_2) \pi/4} \sin \theta \ d\theta  \int_{(1-\epsilon_1) \pi/4}^{(3-\epsilon_1) \pi/4} d\varphi \  
\Psi _{k,\ell,m,n}(r,\alpha,\theta ,\varphi )\Psi_{k^{\prime },\ell^{\prime },m^{\prime},n^{\prime }}(r,\alpha,\theta ,\varphi ) \nonumber\\
 & & =\delta_{k,k^{\prime }}\delta _{\ell,\ell^{\prime }} \delta _{m,m^{\prime }}  \delta _{n,n^{\prime }}N_{k,\ell,m,n} \ \quad \ .
\label{norma}
\end{eqnarray}
Use is made, here,  of the orthogonality properties of 
 Gegenbauer, Jacobi and Laguerre polynomials \cite{Abramowitzbook}. 

The full expression of the  eigenenergies is  expressed by 
\begin{eqnarray}
E_{k,\ell,m,n} &\equiv &E_{k,2 \ell+2m+2n}=2\omega \left( 2k+1+\sqrt{\beta +(2 \ell+2 m+2n+2 a + c + d+3)^{2} } \right) ,  \label{3c-42} \\
k &=&0,1,2,...,\qquad \ell=0,1,2,...\qquad m=0,1,2,...\qquad n=0,1,2,...,\;.  \nonumber
\end{eqnarray}

 For illustration, the equation (\ref{3c-99p}), multiplied by $r^{3/2} (\sin \alpha) \sqrt{\sin \theta}$   reads in Cartesian coordinates, for
the symmetric case  (we remind the reader  that $a=c=\sqrt{1/4+\lambda}$)
\begin{eqnarray}
\Psi(x_1,x_2,x_3,x_4) & \propto &\vert (x_1+x_4-x_2-x_3) (x_1+x_3-x_2-x_4) (x_1+x_2-x_3-x_4) \vert^{a+1/2}
 \nonumber\\
    & \times &  [(x_1-x_2)^2+(x_3-x_4)^2]^n \ C_n^{a+1/2} \left(\frac{2 (x_1-x_2) (x_4-x_3)}{(x_1-x_2)^2+(x_3-x_4)^2} \right) \nonumber\\
          & \times & [ 4 (x_1^2+x_2^2+x_3^2 +x_4^2) -(x_1+x_2+x_3+x_4)^2]^m \nonumber\\
     & \times & P_m^{(2 n + 2 a + 1,a)} \left( \frac{8 (x_1 x_2+x_3 x_4) -  (x_1+x_2+x_3+x_4)^2}{4 (x_1^2+x_2^2+x_3^2
     +x_4^2) -(x_1+x_2+x_3+x_4)^2} \right) \nonumber\\  
         & \times & \vert x_1+x_2+x_3+x_4 \vert^{d+1/2} (x_1^2+x_2^2+x_3^2+x_4^2)^{q} \nonumber\\
      & \times &  P_{\ell}^{(2 m +2 n + 3 a + 2,  d)}\left(
     \frac{2 (x_1+x_2+x_3+x_4)^2-(x_1^2+x_2^2+x_3^2+x_4^2)}{x_1^2+x_2^2+x_3^2+x_4^2} \right) \nonumber\\
     & \times & \exp(- \omega (x_1^2+x_2^2+x_3^2+x_4^2)/2) \ L_k^{p} ( \omega (x_1^2+x_2^2+x_3^2+x_4^2))
\end{eqnarray}
with
\bea 
p  & = & \sqrt{\beta +(  2 \ell + 2 m + 2 n + 3 a + d + 3)^2} \nonumber\\
q & = & p/2- (2 m + 2 n + 3 a + d + 3)/2  \ . \nonumber
\eea 
The eigenvalue Eq.(\ref{3c-42}) shows a degeneracy, i.e., all solutions such that  the equality $\ell+m+n=N$ is satisfied, $N$ being fixed,      
correspond to the same eigenvalue. This imply that any combination of solutions such that $\ell+m+n$ is constant
is also solution. We can then obtain, $k,\ell,m+n$ being fixed, a unique solution, symmetric in the permutation on the set of variables $\{x_1,x_2,x_3,x_4\}$ when $n+m$ is even.  This latter symmetric  solution is in fact proportional to $\sum_{\sigma \in S_4} \Psi(x_{\sigma(1)},x_{\sigma(2)},x_{\sigma(3)},x_{\sigma(4)})$
where $\ S_4$ denotes the permutation group of four elements.

Let us now consider  the irregular  "polynomial" solutions  corresponding to $1/2-a$.  We have to replace $a$ by $-a$ in all equations, from 
Eq.(\ref{phin}) until Eq.(\ref{3c-42}).
Recall that for $-1/4 < \lambda <3/4$, the irregular solutions, Eq.(\ref{phin}), are square integrable, as seen before.
The Sturm-Liouville operator (\ref{3c-7}) is self-adjoint for $\lambda \ne -1/4$.
It has to be added  that, a $\delta$ pathology  occurs in (\ref{3c-99p})  for $a=1/2$ ($\lambda=0$).
We consider values of $\lambda$ in $]-1/4,0[ \cup ]0,3/4[$.

We next examine the impact on the change $a \mapsto -a $ on  the function $\Theta_{m,n}(\theta)$,
Eq.(\ref{thetap}). We remind the reader that $|c|=|a|, a=\pm \sqrt{1/4 + \lambda}$. Here 
$a=-\sqrt{1/4 + \lambda}$. Also, we allow 
 $c=\pm  \sqrt{1/4 + \lambda}$ i.e., we consider negative values of $c$. We have $|a|=|c|$ but 
we allow  $a=\pm c$ (both quantities being allowed to be negative and having a different sign).
Note that another situation corresponds to  $a=\sqrt{1/4 + \lambda}=-c$.
But now we concentrate on   $a=-\sqrt{1/4 + \lambda}, c=\pm a$.
 
First, to ensure the self-adjointness of the Sturm-Liouville operator Eq.(\ref{3c-8}),
we have to impose both $B_n \ne 0$ for every $n$ and $\lambda \ne -1/4$.  Taking into account 
Eq.({\ref{bnp}), for $a$ changed into $-a$, we have clearly to ask for $a \ne 1/2$ which is equivalent to $\lambda \ne 0$.
These latter conditions are fulfilled for  $\lambda$ in $]-1/4,0[ \cup ]0,3/4[$.

Then we examine for which value of $a$ we obtain square integrable solutions
$\Theta_{m,n}(\theta)$  for every value of $n$. Since
\be
\frac{1}{2} +b_n=2 n  + \frac{3}{2}  -2 a \quad (a=\frac{1}{2} \sqrt{1+ 4 \lambda})  \ ,
\label{irr0}
\ee
we have
\be
(\forall n \geq 0)  \qquad\frac{1}{2}+ b_n  \geq \frac{3}{2} - 2 a   \ .
\label{irr1}
\ee
The function $\Theta_{m,n}(\theta)$, Eq.(\ref{thetap}), leads
to square integrable solutions for every value of $n$ if $a<1$, corresponding to $b_n+1>0, \ (\forall n)$. 
This latter inequality happens for  $\lambda <3/4$. This is satisfied when $\lambda \in  ]-1/4,0[ \cup ]0,3/4[$
 (see the paragraph below Eq.(\ref{phin})).
 As far as the term $\lambda/\cos^2 \theta$ is concerned  irregular solutions happen when $c=\sqrt{1/4+ \lambda}$ is changed in $-c$ in Eq.(\ref{thetap}).
These solutions are square integrable when $c < 1$  i.e.,  for $\lambda <3/4$.

Consider the last angular  equation concerning the variable $\alpha$. The differential operator is self-adjoint
 provided  that 
\be
(\forall m \geq 0)  (\forall n \geq 0) \qquad c_{m,n} = 2 m + 2 n + 2  \pm c  - 2 a  \geq  2- 2 a \pm c  \ ,
\label{irr2}
\ee
is non zero.
So that the self-adjointness is ensured when $2- 2 a \pm c >0 $ or equivalently $2- 2 a \pm a >0. $   The quantity $2- a  $ is always positive 
in the domain of acceptable $\lambda$ whereas 
 $2- 3 a $ is positive for  $\lambda < 7/36 \simeq 0.194444$.  Also $\mu$ has to be different from $-1/4$.

The square integrability of the function $G_{\ell,m,n}$, Eq.(\ref{Glmn}), 
is ensured by $c_{m,n}+1 >0,  \ (\forall m) \ ( \forall n)$
i.e. when $3-2 a \pm c  >0$ or equivalently $3-2 a \pm a  >0$ .
Both  quantities $3-2 a \pm  a $  are alway positive for $-1/4 < \lambda <3/4$.
 This defines a domain in $\lambda$ of acceptable solutions.
Also if $d$ is allowed to be negative we must have $-1/4 < d <3/4$.

  As far as the radial equation is concerned, 
the constraint $\beta + D_{\ell,m,n} >0$ allows us to treat the centrifugal barrier in the vicinity of $r=0$. 
(see the above discussion for  $\beta + D_{\ell,m,n} \leq 0$.)
Taking into account the definition of $D_{\ell,m,n}$,  Eq.(\ref{3c-31}), we have
\begin{equation}
\beta +D_{\ell,m,n}=\beta +\left(2 \ell+2 m + 2 n -  2 a \pm c + d + 3 \right)^{2}> 0 , \quad
\forall n\geq 0,\quad \forall m\geq 0, \quad \forall \ell \geq 0.  \label{3c-32bispi}
\end{equation}
This is satisfied   for every $\{\ell,m,n\}$ when $\beta >0$ and  (see Eq.(\ref{3c-32bispi}))
\be
\pm c -2 a + 3+ d > \sqrt{ -\beta }   \qquad\quad \beta \leq 0 
\ee
This condition defines a domain of acceptable values of $\beta$ depending on the values of 
$\lambda$ reminding that $\lambda \in ]-\frac{1}{4},0[ \cup ]0,\frac{3}{4}[$.
  Under such conditions, the radial solutions, Eq.(\ref{3c-37}),  are square integrable because
 $\kappa_{\ell,m,n} >0$.

Note that the spectrum for irregular solutions has eigenvalues lower than the ones corresponding to the  regular solutions. This spectrum is given by

\begin{eqnarray}
E_{k,\ell,m,n}^{(<)} &\equiv &E_{k,\ell+m+n}^{(<)}=2\omega \left( 2k+1+\sqrt{\beta +(2 \ell+2 m+2 n \pm c+d-2 a+3)^{2}} \right) ,  \label{3c-42m} \\
k &=&0,1,2,...,\qquad \ell=0,1,2,...\qquad m=0,1,2,...\qquad n=0,1,2,...,\;.  \nonumber
\end{eqnarray}

Also $d$ can be changed in $-d=-\sqrt{1/4+\mu}$. The requirement 
\bea
\pm c -2 a + 3- d & >& \sqrt{ -\beta }   \qquad\quad \beta  \leq  0 \nonumber\\
     \pm c -2 a + 3- d  & >0 &  \qquad\quad  \beta  \geq  0
\eea
ensures the self-adjointness of Eq.(\ref{3c-10}).

\section{D-dimension}

The generalization to the  $D$-dimensional space follows the same strategy as in section {\bf 2}.
We consider the Hamiltonian:

\begin{equation}
H=\sum_{i=1}^{4}\left( -\Delta_i
+\omega ^{2} \vec{r_{i}}^{2}\right) +2 \lambda  \sum_{i \ne j \ne k \ne m} \frac{1}{
(\vec{r_{i}}+\vec{r_j} - \vec{r_{k}}-\vec{r_m})^2}+\frac{4 \mu }{(\sum_{i=1}^{4}\vec{r_{i}})^2}+\frac{ \beta}{\sum_{i=1}^4
 \vec{r_i}^2}
\label{3c-1d}
\end{equation}

Setting as in section {\bf 2}
\bea
\vec{R} & = & \frac{\vec{r_1}+\vec{r_2}+\vec{r_3}+\vec{r_4}}{2} \nonumber\\
\vec{s} & = & \frac{\vec{r_1}+\vec{r_2}-\vec{r_3}-\vec{r_4}}{2}  \nonumber\\
\vec{t} & = & \frac{\vec{r_1}+\vec{r_3}-\vec{r_2}-\vec{r_4}}{2}  \nonumber\\
\vec{u} & = & \frac{\vec{r_1}+\vec{r_4}-\vec{r_2}-\vec{r_3}}{2} 
\label{coordd}
\eea
the Hamiltonian Eq.(\ref{3c-1d}) becomes 

\begin{eqnarray}
H &=&-\Delta_{\vec{R}} -\Delta_{\vec{s}}-\Delta_{\vec{t}}-\Delta_{\vec{u}}
  +\omega^{2}(R^2+ s^2+ t^2+u^2) 
 \nonumber\\
&& + \frac{ \mu }{R^2} + \lambda \left( \frac{1}{s^{2}}+\frac{1}{t^{2}} +  
\frac{1}{u^{2}}\right) +\frac{\beta}{R^2+ s^2+ t^2+u^2}.
\label{H2d}
\end{eqnarray}

Since the potential does not depend on the angles between $\vec{R},\vec{s},\vec{t},\vec{u}$,
the separation of angular and radial variables, together with the use of the hyperspherical harmonics \cite{erd},
allows us to write

\be  
\Phi(\vec{R},\vec{s},\vec{t},\vec{u})=\frac{\Phi^{(\ell_r,\ell_s,\ell_t,\ell_u)}(R,s,t,u)}{(R s t u)^{(D-1)/2}} \  Y_{\ell_R,[M_R]}(\Omega_R)  \ 
Y_{\ell_s,[M_s]}(\Omega_s) Y_{\ell_t,[M_t]}(\Omega_t) Y_{\ell_u,[M_u]}(\Omega_u) \ .
\ee

Here $[M]$ denotes the set $[M]=\{m_1,m_2,...,m_p\},p=D-2$ satisfying 
$\ell=m_0 \geq m_1 \geq m_2 ... \geq m_p \geq 0$ and  \cite{erd}

\be
Y_{\ell,[M]} = e^{ \pm i m_p \phi} \  \prod_{k=1}^p (\sin{\theta_k})^{m_k} \  
\prod_{k=0}^{p-1} C_{m_k-m_{k+1}}^{m_{k+1}+p/2-k/2} (\cos{\theta_{k+1}})
\ee

where the hyperspherical harmonics $Y_{\ell,[M]}$ are given in terms of the Gegenbauer polynomials $C_n^{a}(x)$.
For $D=2$, $Y_{\ell}=\exp(i \ell \phi), \ell \in {\cal Z}\ $ and $ \phi \in [0,2 \pi]$.
Recall that the hyperspherical polar coordinates are given by

\bea
x_1  & = & r \cos{\theta_1} \nonumber\\
x_2  & = & r \sin{\theta_1} \  \cos{\theta_2} \nonumber\\
x_3  & = & r \sin{\theta_1} \  \sin{\theta_2} \cos{\theta_3} \nonumber\\
   & & ... \nonumber\\
x_{p+1} & = & r \sin{\theta_1} \  \sin{\theta_2}  .. \sin{\theta_p} \cos{\phi} \nonumber\\
 x_{p+2} & = & r \sin{\theta_1} \  \sin{\theta_2}  .. \sin{\theta_p} \sin{\phi}
\eea  

with $\theta_k, \in [0,\pi], k=1,2,..p$  and $\phi \in [0,2 \pi]$.
Setting $(D-3)/2=md$ (here $md \geq -1/2$) we obtain

\bea 
& & \left( -\frac{\partial ^{2}}{\partial R^2}-\frac{\partial ^{2}}{\partial
s^{2}}-\frac{\partial ^{2}}{\partial t^{2}}-\frac{\partial ^{2}}{\partial u^{2}}
  +\omega^{2}(R^2+ s^2+ t^2+u^2)   + \frac{ \mu+ (\ell_R+md) (\ell_R+md+1) }{R^2}      \right.
 \nonumber\\
& & \left.  + \frac{ \lambda+ (\ell_s+md) (\ell_s+md+1) }{s^2} +    \frac{ \lambda+ (\ell_t+md)( \ell_t+md+1) }{t^2} \right. \nonumber\\  
&& \left. +\frac{ \lambda+ (\ell_u+md)( \ell_u+md+1) }{u^2}
 +\frac{\beta}{R^2+ s^2+ t^2+u^2} \right) \Phi^{(\ell_r,\ell_s,\ell_t,\ell_u)}(R,s,t,u)=0
\label{psid}
\eea

We introduce the following  hyperspherical coordinates: 
\begin{eqnarray}
R &=&r\cos \alpha, \quad s=r\sin \alpha \cos \theta, \quad t=r\sin \alpha \sin \theta \sin \varphi,
\quad u=r\sin \alpha \sin \theta \cos \varphi ,\quad  \nonumber \\
0 &\leq &r<\infty ,\quad \quad  0\leq \alpha \leq \frac{\pi}{2} ,\quad\quad\quad\quad   0 \leq \theta \leq \frac{\pi}{2} ,\quad \quad\quad\quad\quad 0\leq
\varphi \leq\frac{\pi}{2}  .  \label{3c-18bisd}
\end{eqnarray}
Here  $\alpha,\theta,\varphi \in [0,\pi/2]$ because $R,s,t,u$ are positive.

The stationary Schr\"{o}dinger equation is then written:

\begin{eqnarray}
&&\left\{ -\frac{\partial ^{2}}{\partial r^{2}}-\frac{3}{r}\frac{\partial }{
\partial r}+\omega ^{2}r^{2}  +\frac{\beta}{r^2} +  \frac{1}{r^{2}}\left[
 -\frac{\partial ^{2}}{\partial \alpha ^{2}}   -2 \cot \alpha \frac{\partial }{\partial \alpha } + \frac{\mu + (\ell_R+md) (\ell_R+md+1) }{\cos^2 \alpha}
 \right.\right. \nonumber\\
&+& \left.\left. 
\frac{1}{\sin ^{2}\alpha}  \left(  -\frac{\partial ^{2}}{\partial \theta ^{2}}-\cot \theta \frac{\partial }{\partial \theta }
+\frac{\lambda +  (\ell_s+md) (\ell_s+md+1) }{ \cos^{2}\theta }
 \right.\right. \right. \nonumber\\
 & & \left.\left.\left. + \frac{1}{\sin^{2}\theta }\left(- 
\frac{\partial ^{2}}{\partial \varphi ^{2}}+ \frac{\lambda + (\ell_t+md)( \ell_t+md+1)}{\sin^{2}\varphi } \right.\right. \right.\right. \nonumber\\
& & \left.\left.\left.\left. +  \frac{\lambda + (\ell_u+md)( \ell_u+md+1)}{\cos^{2}\varphi }
\right) \right) \right] \right\} \Psi(r,\alpha,\theta,\varphi)  =E\Psi(r,\alpha,\theta,\varphi)   ,  \label{3c-49d}
\end{eqnarray}

where $\Psi (r,\alpha,\theta ,\varphi )$ represents the eigensolutions  associated to eigenenergy $E$.

Assuming 
\begin{equation}
\Psi_{k,\ell,m,n} (r,\alpha,\theta ,\varphi )=\frac{F_{k,\ell,m,n}(r)}{r \sqrt{r}} \frac{G_{\ell,m,n}(\alpha)}{\sin \alpha}
 \frac{\Theta_{m,n} (\theta )}{\sqrt{\sin
\theta }}\Phi_n (\varphi )  \label{3c-6d} 
\end{equation}
we obtain
\begin{equation}
\left( -\frac{d^{2}}{d\varphi ^{2}}+ \frac{\lambda + (\ell_t+md)( \ell_t+md+1)}{\sin^{2}\varphi }+
 \frac{\lambda + (\ell_u+md)( \ell_u+md+1)}{\cos^{2}\varphi } 
 \right) \Phi_n (\varphi )=B_n\Phi_n (\varphi ),\quad  \label{3c-7d}
\end{equation}
\begin{equation}
\left( -\frac{d^{2}}{d\theta ^{2}}+\frac{(B_n-\frac{1}{4})}{\sin ^{2}\theta }
+\frac{\lambda + (\ell_s+md)( \ell_s+md+1) }{\cos^2\theta} \right) \Theta_{m,n} (\theta )=C_{m,n} \Theta_{m,n} (\theta ),\qquad  \label{3c-8d}
\end{equation}
\begin{equation}
\left( -\frac{d^{2}}{d\alpha ^{2}}+\frac{C_{m,n}-\frac{1}{4}}{\sin ^{2}\alpha }
+\frac{\mu +  (\ell_R+md)( \ell_R+md+1)}{\cos^2 \alpha} \right) G_{\ell,m,n} (\alpha )=D_{\ell,m,n} G_{\ell,m,n} (\alpha ),\qquad  \label{3c-9d}
\end{equation}
and 
\begin{equation}
\left( -\frac{d^{2}}{dr^{2}}+\omega ^{2}r^{2}+\frac{\beta +D_{\ell,m,n}-\frac{1}{4} }{r^{2}
}\right) F_{k,\ell,m,n}(r)=E_{k,\ell,m,n}\  F_{k,\ell,m,n}(r) \ .  \label{3c-10d}
\end{equation}
The equation (\ref{3c-7d}) is solved in the interval $[0,\pi/2]$.
The regular eigensolutions of equation (\ref{3c-7d}) on $[0,\pi/2]$ read  \cite{Perelomov1983,JZ}
\be     
 \Phi_n (\varphi )= (\sin \varphi)^{1/2+a} (\cos \varphi)^{1/2+b}  P_n^{(a,b)}(\cos2 \varphi) 
\label{phind}
\ee
where the  $ P_n^{(a,b)}$ denote the Jacobi Polynomials \cite{erd}. The  corresponding  eigenvalues 
are
\be  
B_n=\left(a+b+1 + 2 n \right)^2 
\label{bnd}
\ee
with 
\be  
 n=0,1,2,...  \qquad\quad a=\sqrt{\lambda+  (\ell_t+md+1/2)^2} \qquad\quad b=\sqrt{\lambda+  (\ell_u+md+1/2)^2}
\label{symbad}
\ee 
The operator in Eq.(\ref{3c-7d}) is self-adjoint for every $\ell_t,\ell_u$ provided that $\lambda+(md+1/2)^2 > 0$.
Introducing the parameter 
\be   
 b_n=\sqrt{B_n}=1+ a +b + 2 n
\label{bnpd}
\ee
(where  we have only considered the positive root $b_n=\sqrt{B_n}$ because the other root $b_n=-\sqrt{B_n}$
leads to non-square integrable solutions for most values of $n$) 
the solution of Eq.(\ref{3c-8d}) on the interval $[0,\pi/2]$ can be obtained as  \cite{Perelomov1983,BLL2009}
\bea   
 \Theta_{m,n}(\theta)&=& (\sin \theta)^{b_n+1/2} \ (\cos  \theta)^{c+1/2} \ P_m^{(b_n,c)}(\cos 2 \theta)
\label{thetapd} \\
  & & m=0,1,2,... \qquad\quad c =\sqrt{ \lambda+(\ell_s+md+1/2)^2} \nonumber\\
 \nonumber
\eea
corresponding to the eigenvalue
\be   
C_{m,n}= (2 m + b_n + c + 1)^2  \ .
\label{Cmnd}
\ee
In Eq.(\ref{thetapd}) the $P_m^{(b_n,c)}$ denote the Jacobi polynomials \cite{erd}.
The operator in Eq.(\ref{3c-8d}) is self-adjoint for every $\ell_t,\ell_u,\ell_s$ provided that  $\lambda+(md+1/2)^2 > 0$.
The second condition namely  $B_n \ne 0$ i.e.  $2 \sqrt{\lambda+ (md+1/2)^2} +1 >0$ is always satisfied.

The equation (\ref{3c-9d}) is solved in the same manner as done for equation  (\ref{3c-8d}).
Setting 
\be  
c_{m,n}=\sqrt{C_{m,n} \ }=2 m + b_n + c + 1
\label{cmnd}
\ee
(where  we have only considered the positive root $c_{m,n}=\sqrt{C_{m,n}}$ because the other root $c_{m,n}=-\sqrt{C_{m,n}} $
leads to non-integrable solutions for most values of $m$)
the solution writes as
\bea 
 G_{\ell,m,n} (\alpha ) & = &  (\sin \alpha)^{c_{m,n}+1/2} \ ( \cos  \alpha)^{d+1/2} \
  P_{\ell}^{(c_{m,n},d)}(\cos 2 \alpha) \label{Glmnd} \\
   & & \ell=0,1,2,... \qquad\quad d =\sqrt{ \mu+(\ell_R+md+1/2)^2}  \label{ddd} 
\eea
corresponding to the eigenvalue
\be   
D_{\ell,m,n} = (2 \ell +c_{m,n}  +d + 1)^2  \ .
\label{3c-31d}
\ee 
The operator in Eq.(\ref{3c-9d}) is self-adjoint for every $\ell_t,\ell_u,\ell_s,\ell_R$ provided that $\lambda+(md+1/2)^2 > 0$.
The condition $C_{m,n} \ne 0$ i.e.  $3 \sqrt{\lambda+ (md+1/2)^2} +2 >0$ is always satisfied.

On the other hand, the reduced radial equation (\ref{3c-10d})
is solved in the interval  $0\leq r<\infty ,$ with the condition of square integrability for the solutions.
It implies $F_{k,\ell,m,n}(r) \to 0$ as $r \to \infty$.
We have to impose $\beta + D_{\ell,m,n} > 0 $ in order to treat the centrifugal barrier in the vicinity of $r=0$.
\begin{equation}
\beta +D_{\ell,m,n}=\beta +\left(2 \ell+2 m + 2 n +  a +b + c + d + 3 \right)^{2}> 0 , \quad
\forall n\geq 0,\quad \forall m\geq 0, \quad \forall \ell \geq 0.  \label{3c-32bispd}
\end{equation}
For positive $a,b,c,d$ the  quantity $\beta +D_{\ell,m,n}$ is minimal for  $n=0,m=0,\ell=0$ and $a=b=c=d=0$ ( recall that $a,b \geq 0$, see (\ref{symbad}),
 that $c \geq 0$, see (\ref{thetapd}) and that $d \geq 0$  see (\ref{ddd}))
It put constraint on  $\beta$ to satisfy  $\beta >-9$ when $a,b$,$c$ and $d$ are positive.
 We introduce the auxiliary parameter  $\kappa _{\ell,m,n}$ defined by
\begin{equation}
\kappa _{\ell,m,n}^{2}=\beta +D_{\ell,m,n}, \qquad \kappa _{\ell,m,n}=
\sqrt{\beta +D_{\ell,m,n} \ }.
\label{3c-33d}
\end{equation}
 The solution of the radial equation (\ref{3c-10d})  is \cite{BLL2009} 
\begin{equation}
F_{k,\ell,m,n}(r)=r^{\kappa _{\ell,m,n}+\frac{1}{2}}\exp \left(-\frac{\omega r^{2}}{2}
\right)L_{k}^{(\kappa _{\ell,m,n})}(\omega r^{2}),\qquad k=0,1,2...,  \label{3c-37d}
\end{equation}
and it is associated to the eigenenergy
\begin{equation}
E_{k,\ell,m,n}=2\omega (2k+\kappa _{\ell,m,n}+1),\qquad k=0,1,2....  \label{3c-38d}
\end{equation}
The $L_k^{(q)}$ are the generalized Laguerre polynomials \cite{erd}.

Taking into account all  information, we conclude that the physically acceptable solutions of 
the Schr\"odinger equation  (\ref{3c-49d}) are
\begin{eqnarray}
\Psi _{k,\ell,m,n}(r,\alpha,\theta ,\varphi ) &=&r^{\sqrt{\beta +(2 \ell+ 2m +2n+a+b+ c+d+3)^{2} \ }-
1}e^{-\frac{\omega r^{2}}{2}}L_{k}^{\left( \sqrt{\beta +(2 \ell+ 2m +2n+a+b+c+d+3)^{2} \ }
\right) }(\omega r^{2})  \nonumber \\
&&\times (\sin \alpha )^{2m +2n+ a+b+c+\frac{5}{2}} \ ( \cos \alpha )^{d+1/2} P_{\ell}^{(2 m + 2 n + a+b+ c +2,d)}(\cos 2 \alpha)
  \nonumber \\
&&\times (\sin \theta )^{2 n+a+b+\frac{3}{2}} ( \cos \theta)^{c+\frac{1}{2}} 
P_{m}^{\left( 2n +  a +b+ 1,c \right)}(\cos 2 \theta )  \nonumber \\
&&\times ( \sin \varphi )^{a+\frac{1}{2}}  ( \cos \varphi )^{b+\frac{1}{2}} P_{n}^{( a,b)}(\cos 2 \varphi ),\,  \label{3c-39ppd} \\
k &=&0,1,2,...,\qquad \ell=0,1,2,...,\qquad m=0,1,2,...,,\qquad n=0,1,2,...,   \nonumber
\end{eqnarray}

The normalization constants $N_{k,\ell,m,n}$ can be  calculated from
\begin{eqnarray}
&& \int_{0}^{+\infty }r^{3}dr\int_{0}^{\pi/2 }\sin^2(\alpha)d\alpha 
\int_{0}^{\pi/2} \sin \theta \ d\theta \int_0^{\frac{\pi }{2}}d\varphi \ 
\Psi _{k,\ell,m,n}(r,\alpha,\theta ,\varphi )\Psi_{k^{\prime },\ell^{\prime },m^{\prime},n^{\prime }}(r,\alpha,\theta ,\varphi ) \nonumber\\
 & & =\delta_{k,k^{\prime }}\delta _{\ell,\ell^{\prime }} \delta _{m,m^{\prime }}  \delta _{n,n^{\prime }}N_{k,\ell,m,n},
\label{normad}
\end{eqnarray}
use is made, here,  of the orthogonality properties of 
 Gegenbauer, Jacobi and Laguerre polynomials \cite{Abramowitzbook}. 

The full expression of the  eigenenergies is  expressed by 
\begin{eqnarray}
E_{k,\ell,m,n} &\equiv &E_{k,\ell+m+n}=2\omega \left( 2k+1+\sqrt{\beta +(2 \ell+2 m+2n+ a +b+ c + d+3)^{2} } \right) ,  \label{3c-42d} \\
k &=&0,1,2,...,\qquad \ell=0,1,2,...\qquad m=0,1,2,...\qquad n=0,1,2,...,\;.  \nonumber
\end{eqnarray}

\section{Conclusions}

 In this paper, we have proposed and solved  a four-body quantum problem
of Wolfes type, in the $D=1$ dimensional  space. The operator we consider
is composed of  the harmonic trap with four-body translationally invariant
potential
proposed by Wolfes with additional non-translationally invariant four-body
potentials.
We explicitly give the full solutions of the corresponding Schr\"odinger
equation,
 namely the wave functions in terms of the angular and radial variables together
with
 the energy spectrum. \\
We have been able, also, to generalize and solve this  problem to any 
$D$-dimensional space,
where the explicit eigensolutions and the corresponding spectrum have been
calculated. \\
We also investigate the domain of coupling constant for which  the irregular
solutions
 become square integrable, and physically acceptable. \\

 As a perspective , we can note that this proposed exactly solved four-body
problem can be transformed  to the case where the harmonic trap is replaced  by
any attractive hypercentral potential, depending on the hyperradius
$r=\sqrt{\sum_{i=1}^4 x_i^2}$. This was done
 for the hypercentral    "Coulomb" type potential for the three-body problem,
giving rise to both discrete and continuous spectra   \cite{KB}.   

{\bf Acknowledgements}
We thank Dr. R.J. Lombard for interesting discussions. 
One of us (A.B.) is very grateful to the Theory Group of the IPN 
Orsay for its kind hospitality.

\end{document}